\documentclass[12pt]{article}
\usepackage{amsmath}
\usepackage{amssymb}
\usepackage[final]{graphicx}
\usepackage{enumerate}
\usepackage{natbib}
\usepackage{url} 

\usepackage{float}
\usepackage{chngcntr}
\usepackage{tocloft}

\usepackage{amsthm}

\theoremstyle{definition}

\usepackage{mathptmx}

\DeclareMathOperator{\vect}{vec}

\usepackage{lscape}
\usepackage{ltxtable}
\usepackage{array,multirow,makecell}

\usepackage{graphicx}
\usepackage{amsmath}
\usepackage{float}
\usepackage{array,multirow,makecell}
\usepackage{lscape}
\usepackage{tabularx}

\newcommand{\blind}{1}

\begin{document}





\if1\blind
{
  \title{\bf Time, Frequency \& Time-Varying Causality Measures in Neuroscience}
  \author{Sezen Cekic
   \hspace{.2cm}\\
    Methodology and Data Analysis, Department of Psychology, \\
    University of Geneva,\\
    Didier Grandjean\\
    Neuroscience of Emotion and Affective Dynamics Lab, \\Department of Psychology, \\ University of Geneva,\\
	and \\
    Olivier Renaud \\
    Methodology and Data Analysis, Department of Psychology, \\ University of Geneva}
  \maketitle
} 

\fi

\if0\blind
{
  \title{\bf Time, Frequency \& Time-Varying Causality Measures in Neuroscience}
  \author{}
   \maketitle

} 

\fi

\bigskip
\begin{abstract}
This article proposes a systematic methodological review and objective criticism of existing methods enabling the derivation of time-varying Granger-causality statistics in neuroscience. The increasing interest and the huge number of publications related to this topic calls for this systematic review which describes the very complex methodological aspects. The capacity to describe the causal links between signals recorded at different brain locations during a neuroscience experiment is of primary interest for neuroscientists, who often have very precise prior hypotheses about the relationships between recorded brain signals that arise at a specific time and in a specific frequency band. The ability to compute a time-varying frequency-specific causality statistic is therefore essential. Two steps are necessary to achieve this: the first consists of finding a statistic that can be interpreted and that directly answers the question of interest. The second concerns the model that underlies the causality statistic and that has this time-frequency specific causality interpretation. In this article, we will review Granger-causality statistics with their spectral and time-varying extensions. 
\end{abstract}

\noindent%
\vfill

\newpage

\maketitle

\section{Introduction} 
\label{Introduction_LITT}

The investigation of the dynamical causal relationships between neuronal populations is a very important step towards the overall goal of understanding the links between functional cerebral aspects and their underlying brain mechanisms. This investigation requires statistical methods able to capture not only functional connectivities (e.g., symmetrical relationships), but also, and probably more importantly, effective connectivities (e.g., directional or causal relationships) between brain activities recorded during a specific task or stimuli exposure.

The question of how to formalize and test causality is a fundamental and philosophical problem. A statistical answer, which relies on passing from causality to predictability, was provided in the 1960's by the economist Clive Granger and is known as ``Granger causality''. According to \cite{granger_investigating_1969}, if a signal $X$ is causal for another signal $Y$ in the Granger sense, then the history of $X$ should contain information that helps to predict $Y$ above and beyond the information contained in the history of $Y$ alone. It is the axiomatic imposition of a temporal ordering that allows us to interpret such dependence as causal: \textit{``The arrow of time imposes the structure necessary''} (\cite[p.~139]{granger_testing_1980}). The presence of this relation between $X$ and $Y$ will be referred to ``Granger causality'' throughout the text.

\cite{granger_investigating_1969} adapted the definition of causality proposed by \cite{e_theory_1956} into a practical form and, since then, Granger causality has been widely used in economics and econometrics. It is however only during the past few years that it has become popular in neuroscience (see \cite{pereda_nonlinear_2005} and \cite{bressler_wienergranger_2011} for a review of Granger causality applied to neural data).

Since its causal nature relies on prediction, Granger causality does not necessarily mean ``true causality''. If the two studied processes are jointly driven by a third one, one might reject the null hypothesis of non-Granger causality between signals although manipulation of one of them would not change the other, which contradicts what ``true causality'' would have implied.

Granger causality may also produce misleading results when the true causal relationship involves more variables than those that have been selected and so the accuracy of its causal interpretation relies on a suitable preliminary variable selection procedure (\cite{pearl_causal_2009}).

If we concentrate on just two signals, the problem is twofold: the first part is the choice of a suitable causality statistic that can easily be interpreted and that answers the question of interest. This said, the statistic needs to rely on a model which intrinsically includes this prediction or Granger-causality principle, and so the second part of the problem is to define and properly estimate this fundamental statistical model. A wrong statistical model indeed may lead to a wrong causality inference.

The scope of this article is to review and describe existing Granger-causality statistics in the time and frequency domains and then to focus on their time-varying extensions. We will describe existing estimation methods for time-varying Granger-causality statistics, in order to give the reader a global overview and some insight on the pertinence of using a given method depending on the research question and the nature of the data.

In Sections \ref{Time domain causality} and \ref{Frequency domain causality}, we will present time and frequency-domain Granger-causality statistics in the stationary case. In Section \ref{Time-varying Granger causality}, we will discuss their time-varying extensions in terms of time-varying causal model estimation. In Section \ref{Existing toolboxes}, we will outline existing toolboxes allowing us to derive time-varying frequency-specific Granger-causality statistics and then discuss the limitations and the potential application of these statistics in neuroscience in Section \ref{Discussion}.

To our knowledge, there is no systematic methodological review and objective criticism of existing methods that lead to time-varying Granger-causality statistics. The increasing interest reflected by the number of publications related to this topic in neuroscience justifies this literature review undertaken from a statistical viewpoint. 

\section{Stationarity}
Many Granger-causality models rely on the assumption that the system analyzed is covariance stationary. Covariance stationarity (also known as weak- or wide-sense stationarity) requires that the first moment and the covariance of the system do not vary with respect to time.

A random process $\mathbf{Z}_t$ is covariance stationary if it satisfies the following restrictions on its mean function:
\begin{equation}
    \mathbb{E}[\mathbf{Z}(t)] = m_{\mathbf{Z}}, \qquad \forall t \in \mathbb{R},
\end{equation}
and on its autocovariance function:
\begin{equation}
\begin{aligned}
 &\mathbb{E}[(\mathbf{Z}(t_1)-m_{\mathbf{Z}})(\mathbf{Z}(t_2)-m_{\mathbf{Z}})] = 
 C_{\mathbf{Z}}(t_1, t_2) = C_x(\tau) \,\! ,\mbox{ where }  \tau = t_1
 - t_2, \quad \forall t_1, t_2 \in \mathbb{R}.
\end{aligned}
\end{equation}
The first property implies that the mean function $m_{\mathbf{Z}}$ is constant with respect to $t$. The second property implies that the covariance function depends only on the difference between $t_1$ and $t_2$. The variance is consequently constant as well.

\label{Stationarity}

\section{Time Domain Causality}
\label{Time domain causality}

We will first discuss the simplest case of Granger causality which is defined in the time domain. It is important to note that it requires that the data are stationary.

As mentioned in the introduction, Granger causality is based on prediction and its fundamental axiom is that ``the past and present may cause the future but the future cannot cause the past'' (\cite{granger_investigating_1969}). The origin of Granger-no-causality was stated by Wiener in 1956 and then adapted and defined into practical form by Granger. As we will see,
Granger restates Wiener's principle in the context of autoregressive models (\cite{granger_investigating_1969}). In particular, the main idea lies in the fact that if a signal $X$ is causal for another signal $Y$ in the Granger sense, then past values of $X$ should contain information that helps to predict $Y$ better than merely using the information contained in past values of $Y$ (\cite{granger_investigating_1969}). 

This concept of predicting better with an additional variable can be linked to significance tests in multiple linear regression, where an independent variable is declared significant if the full model explains (predicts) the dependent variable better than the model that does not contain this variable. In many fields these tests are called marginal and are linked to the so-called ``Type III sum of squares'' in ANOVA.

The general criterion of causality is: if the prediction error of a first series given its own past is significantly bigger than its prediction error given its own past plus the past of a second series, then this second series causes the first, in the Granger sense (\cite{granger_testing_1980,granger_investigating_1969,ding_granger_2006}). 

As \cite{chamberlain_general_1982}, \cite{florens_technical_2003} and \cite{chicharro_spectral_2011} point out, the most general criterion of Granger non-causality can be defined based on the equivalence of two conditional densities:
\begin{equation}
f_t(Y_{t}|Y_{t-1}^{t-p}) =f_t(Y_{t}|Y_{t-1}^{t-p},X_{t-1}^{t-p}),
\label{HYP.def}
\end{equation}
where $X_t$ and $Y_t$ are the two recorded time series, $Y_{t-1}^{t-p}$ and $X_{t-1}^{t-p}$ denote the history from time $t-1$ to $t-p$ of $Y$ and $X$ respectively (i.e. [$Y_{t-1},\dots,Y_{t-p}$], and [$X_{t-1},\dots,X_{t-p}$]), and $p$ is a suitable model order. This general criterion is expressed in terms of the distributions only, so it does not rely on any model assumptions (\cite{kursteiner}). Note that in this general definition, $f_t(.)$ can be different for each time, and therefore the general criterion in equation \eqref{HYP.def} includes nonstationary models.

Any existing method for assessing Granger causality can be viewed as a restricted estimation procedure allowing us to estimate the two densities in equation \eqref{HYP.def} and to derive a causality statistic in order to test their difference.

For linear Gaussian autoregressive models, the assumptions are
Gaussianity, homoscedasticity and linearity, which implies
stationarity in most cases. The quantities in equation \eqref{HYP.def} become an autoregressive model of order $p$ (AR($p$)) for the left-hand side:
\begin{equation}
f_t(Y_{t}|Y_{t-1}^{t-p}) = \phi(Y_{t}; \mu=\displaystyle\sum\limits_{j=1}^p \vartheta_{1(j)} Y_{t-j}, \sigma^2=\Sigma_{1}),
\label{AR.def}
\end{equation}
and a vector autoregressive model of order $p$ ({VAR}($p$)) for the right-hand side:
\begin{equation}
f_t(Y_{t}|Y_{t-1}^{t-p},X_{t-1}^{t-p})  =  \phi(Y_{t}; \mu=\displaystyle\sum\limits_{j=1}^p
\vartheta_{11(j)} Y_{t-j}+\displaystyle\sum\limits_{j=1}^p \vartheta_{12(j)} X_{t-j}, \sigma^2=\Sigma_{2}),
\label{VAR1.def}
\end{equation}
where $\phi$ stands for the Gaussian probability density function.

In the next sections, we will present the two widely used approaches for testing hypotheses~\eqref{HYP.def} in the linear Gaussian context. The first one is based on an $F$ statistic expressed as the ratio of the residual variances of models on equations~\eqref{AR.def} and \eqref{VAR1.def} (\cite{geweke_measurement_1982}). The second one is based on a Wald statistic and tests the significance of the causal VAR coefficients (\cite{hamilton_time_1994,lutkepohl2005new}). 

\subsection{Granger-causality criterion based on variances} 
\label{Granger causality criterion 1}
The original formulation of Granger causality (\cite{granger_investigating_1969}) is expressed in terms of comparing the innovation variances of the whole (equation~\eqref{VAR1.def}) and the restricted (equation~\eqref{AR.def}) linear Gaussian autoregressive models (\cite{geweke_measurement_1982,ding_granger_2006}). \cite{granger_investigating_1969} proposed the following quantity to quantify this variance comparison: 
\begin{equation}
F_{X \rightarrow{}Y} = \ln (\dfrac{\Sigma_{1}}{\Sigma_{2}}).
\label{F.def}
\end{equation}
In \cite{hesse_use_2003} and \cite{goebel_investigating_2003} this quantity is estimated by replacing the two variances by estimates. A test based on resampling this statistic is used for assessing the significance.

Geweke made several other important statements for \eqref{F.def} (\cite{geweke_inference_1984,geweke_measurement_1982}). He showed first that the total interdependence between two variables can be decomposed in terms of their two reciprocal causalities plus an instantaneous feedback term. Secondly, he showed that under fairly general conditions, $F_{X \rightarrow{}Y}$ can be decomposed additively by frequency (see Section \ref{Frequency domain causality}). Lastly, he pointed out that it is possible to extend Granger causality to include other series. Based on the conditional densities, the null hypothesis would write
\begin{equation}
f_t(Y_{t}|Y_{t-1}^{t-p},\textbf{W}_{t-1}^{t-p}) =f_t(Y_{t}|Y_{t-1}^{t-p},X_{t-1}^{t-p},\textbf{W}_{t-1}^{t-p}),
\label{multiple_GC.def}
\end{equation}
where $\textbf{W}_{t-1}^{t-p}$ represents a set of variables that are controlled for when assessing the causality from $X$ to $Y$. In the literature, this extension bears the name conditional Granger causality (\cite{ding_granger_2006}).

As explained in \cite{bressler_wienergranger_2011} and \cite{geweke_measurement_1982}, comparing the innovation variances of the whole and restricted linear Gaussian autoregressive models amounts to evaluating the hypothesis
\begin{equation}
\text{Ho:} \quad \Sigma_{1}=\Sigma_{2},
\label{H0_SIGMA}
\end{equation}
which can be assessed through the statistic
\begin{equation}
F = \dfrac{\dfrac{RSS_{r}-RSS_{ur}}{m}}{\dfrac{RSS_{ur}}{T-2m-1}}.
\label{Fstat.def}
\end{equation}
$RSS_{r}$ and $RSS_{ur}$ are the residual sum of squares of the linear models in equations \eqref{AR.def} and \eqref{VAR1.def}, needed to estimate $\Sigma_{1}$ and $\Sigma_{2}$ respectively, and $T$ is the total number of observations used to estimate the unrestricted model.

This statistic follows approximately an $F$ distribution with degrees of freedom $m$ and $T-2m-1$. A significant $F$ may reasonably be interpreted as an indication that the unrestricted model provides a better prediction than does the restricted one, and so that $X$ causes $Y$ in the Granger sense. 

\subsection{Granger-causality criterion based on coefficients} 
\label{Granger causality criterion based on coefficients}

Another way to test for causality between two series under the same conditions as in Section \ref{Granger causality criterion 1} is to estimate model \eqref{VAR1.def} only and to directly test the significance of the VAR coefficients of interest (\cite{hamilton_time_1994,lutkepohl2005new}).
Let us first define the complementary equation of equation \eqref{VAR1.def}
\begin{equation}
f_t(X_{t}|X_{t-1}^{t-p},Y_{t-1}^{t-p})  =  \phi(X_{t}; \mu=\displaystyle\sum\limits_{j=1}^p
\vartheta_{22(j)} X_{t-j}+\displaystyle\sum\limits_{j=1}^p \vartheta_{21(j)} Y_{t-j}, \sigma^2=\Sigma_{3}),
\label{VAR2.def}
\end{equation}
and the variance-covariance matrix of the whole system
\begin{equation}
\Sigma= \left(
\begin{array}{cc}
\Sigma_{2} & \Gamma_{23} \\
\Gamma_{23} & \Sigma_{3} \\
\end{array}
\right),
\label{SIGMA}
\end{equation}
where the off-diagonal elements may or may not be equal to zero.
Testing whether $X$ causes $Y$ in the Granger sense amounts to testing the hypotheses
\begin{equation}
\vartheta_{12(1)}=\vartheta_{12(2)}=\vartheta_{12(3)}=\cdots=\vartheta_{12(p)}=0,
\label{GC2a.def}
\end{equation}
and testing whether $Y$ causes $X$ in the Granger sense amounts to testing 
\begin{equation}
\vartheta_{21(1)}=\vartheta_{21(2)}=\vartheta_{21(3)}=\cdots=\vartheta_{21(p)}=0.
\label{GC2b.def}
\end{equation}
In the context of linear Gaussian autoregressive models, the two null hypotheses \eqref{H0_SIGMA} and \eqref{GC2a.def} are equivalent.

We can observe that the approach using hypothesis \eqref{H0_SIGMA} requires the computation of two models (an AR model and a VAR model), whereas a single VAR model is sufficient for the approach using hypothesis \eqref{GC2a.def}. 

Under joint normality and finite variance-covariance assumptions, the Wald statistic is defined as
\begin{equation}
W = ( \mathbf{\hat{\boldsymbol{\vartheta}}_{12}})' \big( \text{var}(\mathbf{\hat{\boldsymbol{\vartheta}}_{12}})\big )^{-1} ( \mathbf{\hat{\boldsymbol{\vartheta}}_{12}}),
\label{WALD.def}
\end{equation}
where $\mathbf{\boldsymbol{\vartheta}_{12}}$ contains all the parameters $\vartheta_{12(j)}$, for $j=1,\dots,p$. As $T$ increases, this statistic asymptotically follows a $\chi^{2}$ distribution with $p$ degrees of freedom (\cite{lutkepohl2005new}). A significant Wald statistic suggests that at least one of the causal coefficients is different from zero, and, in that sense, that $X$ is causal for $Y$ in the Granger sense. See \cite{sato_method_2006} for an example of application of this statistic in neuroscience.

The time-domain Granger-causality statistics in equations \eqref{Fstat.def} and \eqref{WALD.def} are derived from AR and VAR modelling of the data.
Their relevance therefore relies on the quality of the fitted models. The first issue is the selection of the model order $p$.  Traditional criteria used in time series are the Akaike information criterion (\cite{akaike_new_1974}) and the Bayesian information criterion (\cite{schwarz_estimating_1978}). For the first statistic, in equation \eqref{Fstat.def}, it is advisable to select the same $p$ for the two models.
The second issue is probably often overlooked but of utmost importance. In practice, and particularly for neuroscience data, the plausibility of the assumptions behind these models must be checked before interpreting the resulting tests. This includes analysis of the residuals from the fitted model.

\subsection{Transfer entropy} 
Transfer entropy (TE) is a functional statistic developed in information theory (\cite{schreiber_measuring_2000}). It can be used to test the null hypothesis \eqref{HYP.def} in terms of the distributions themselves, and thus does not rely on the linear Gaussian assumption. It is defined as the Kullback--Leibler distance between the two distributions $f(Y_{t}|Y_{t-1}^{t-p})$  and $f(Y_{t}|Y_{t-1}^{t-p},X_{t-1}^{t-p})$:
\begin{equation}
\begin{split} 
T_{X \rightarrow{}Y}&= \int\cdots\int\!f(y_{t}|y_{t-1}^{t-p},x_{t-1}^{t-p})\ln\dfrac{f(y_{t}|y_{t-1}^{t-p},x_{t-1}^{t-p})}{f(y_{t}|y_{t-1}^{t-p})} \, \mathrm{d}y_{t}\mathrm{d}y_{t-1}^{t-p}\mathrm{d}x_{t-1}^{t-p}\\
&= KL\left\lbrace f(y_{t}|y_{t-1}^{t-p})\parallel f(y_{t}|y_{t-1}^{t-p},x_{t-1}^{t-p})\right\rbrace,
\end{split}
\label{TR.def}
\end{equation}
where the integrals over $y_{t-1}^{t-p}$ and $x_{t-1}^{t-p}$ are both of dimension $p$, and so the overall integral in equation \eqref{TR.def} is of dimension $\left\lbrace 2p+1 \right\rbrace$.

An even more general definition would allow the distributions $f(.)$ to depend on time, letting the transfer-entropy statistic be time-dependent.

It has been shown that for stationary linear Gaussian autoregressive models \eqref{AR.def} and \eqref{VAR1.def}, the indices \eqref{TR.def} and \eqref{F.def} are equivalent (\cite{barnett_granger_2009,chicharro_spectral_2011}).

In its general form, TE is a functional statistic, free from any parametric assumption on the two densities $f(Y_{t}|Y_{t-1}^{t-p})$ and $f(Y_{t}|Y_{t-1}^{t-p},X_{t-1}^{t-p})$. See for example \cite{chavez_statistical_2003},\cite{garofalo_evaluation_2009}, \cite{vicente_transfer_2011}, \cite{wibral_transfer_2011}, \cite{lizier_multivariate_2011}, \cite{b_extraction_2011} and \cite{besserve_causal_2010} for applications of TE in neuroscience. 
Difficulties arise when trying to estimate and compute the joint and marginal densities in equation \eqref{TR.def}. In principle, there are several ways to estimate these two quantities non-parametrically, but the performance of each strongly depends on the characteristics of the data. For a general review of non-parametric estimation methods in information theory see \cite{vicente_transfer_2011} and \cite{hlavavckova2007causality}. For simple discrete processes, the probabilities can be determined by computing the frequencies of occurrence of different states. For continuous processes, which are those of interest for neuroscience, it is more delicate to find a reliable non-parametric density estimation. Kernel-based estimation is among the most popular methods; see for example \cite{victor_binless_2002}, \cite{kaiser_information_2002}, \cite{schreiber_measuring_2000} and \cite{vicente_transfer_2011}.
The major limitation of non-parametric estimation is due to the dimension and the related computational cost. In the present case, the estimation of $f(Y_{t}|Y_{t-1}^{t-p})$ and $f(Y_{t}|Y_{t-1}^{t-p},X_{t-1}^{t-p})$ presents two major limitations due to the curse of dimensionality induced by the model order $p$: a computational limitation, as it implies integration in dimension $2p+1$ in equation \eqref{TR.def}, and the huge number of observations required to non-parametrically estimate the densities, as this number grows exponentially with the dimension. 
Typically, \cite{schreiber_measuring_2000} proposes to choose the minimal $p$, meaning $p=1$, for computational reasons (\cite[~p.462]{schreiber_measuring_2000}).

A toolbox named TRENTOOL provides the computation of TE and the estimation of $f(Y_{t}|Y_{t-1}^{t-p})$  and $f(Y_{t}|Y_{t-1}^{t-p},X_{t-1}^{t-p})$ through kernel estimation (\cite{lindner_trentool:_2011}).
This toolbox enables us to estimate a supplementary parameter, called the embedding delay ($\tau$), which represents the lag in time between each observation of the past values of variables $X$ and $Y$. Equation \eqref{TR.def} then becomes
\begin{equation}
T_{X \rightarrow{}Y}= \int\cdots\int\!f(y_{t}|y_{t-1\tau}^{t-p\tau},x_{t-1\tau}^{t-p\tau})\ln\dfrac{f(y_{t}|y_{t-1\tau}^{t-p\tau},x_{t-1\tau}^{t-p\tau})}{f(y_{t}|y_{t-1\tau}^{t-p\tau})} \, \mathrm{d}y_{t}\mathrm{d}y_{t-1\tau}^{t-p\tau}\mathrm{d}x_{t-1\tau}^{t-p\tau}.\\
\label{TR_tau.def}
\end{equation}
The model order $p$ (called the embedding dimension in this context) is optimized simultaneously with the embedding delay $\tau$ through two implemented criteria. The first is the ``Cao criterion'' (\cite{cao_practical_1997}), which selects $\tau$ on an ``ad hoc'' basis and $p$ through a false neighbour criterion (\cite{lindner_trentool:_2011}). The second is the ``Ragwitz criterion'' (\cite{schreiber_measuring_2000}), which selects $\tau$ and $p$ simultaneously by minimising the prediction error of a local predictor. As discussed in \cite{lindner_trentool:_2011}, the choice of the order $p$ and of the embedding delay $\tau$ is quite important. Indeed, if $p$ is chosen too small, the causal structure may not be captured and thus the TE statistic will be incorrect. On the other hand, using an embedding dimension which is higher than necessary will lead to an increase of variability in the estimation, in addition to a considerable increase in computation time. Typically, \cite{wibral_transfer_2011} select the value of $p$ as the maximum
determined by the Cao criterion from $p=1$ to $4$, and choose the value of $\tau$ following a popular ad hoc option as the first zero of the autocorrelation function of the signal.

TRENTOOL allows us to compute the distribution of the transfer entropy statistic  under the null hypothesis through a permutation method. The data are shuffled in order to break the links between the signals and then the transfer entropy statistic is recomputed on each surrogate dataset (e.g., \cite{wibral_transfer_2011} use $1.9 \times 10^5$ permutations for assessing the significance of the TE statistic). Analyses with TRENTOOL are limited so far to bivariate systems.

The formulation of causality based on the conditional independence in equation \eqref{HYP.def} was later used and theoretically refined in \cite{chamberlain_general_1982} and \cite{florens_technical_2003}. Although less general, the statistics given in equations \eqref{F.def} and \eqref{WALD.def} are much easier to implement and are testable. This probably explains why they have received considerably more attention in applied work.

\section{Frequency Domain Causality}
\label{Frequency domain causality}

\subsection{Geweke--Granger-causality statistic}
As mentioned in Section~\ref{Granger causality criterion 1}, an important advance in  developing the Granger-causality methodology was to provide a spectral decomposition of the time-domain statistics (\cite{geweke_measurement_1982,geweke_measures_1984}).

For completeness, we give below the mathematical details of this derivation. The Fourier transform of equations \eqref{VAR1.def} and \eqref{VAR2.def} for a given frequency $\omega$  (expressed as a system of equations) is
\begin{equation}
\left(
\begin{array}{cc}
\vartheta_{11}(\omega) & \vartheta_{12}(\omega) \\
\vartheta_{21}(\omega) & \vartheta_{22}(\omega) \\
\end{array}
\right)
\left(
\begin{array}{c}
Y(\omega) \\
X(\omega) \\
\end{array}
\right)
=\left(
\begin{array}{c}
\varepsilon_{1}(\omega)\\
\varepsilon_{2}(\omega)\\
\end{array}
\right),
\label{GC_frq1}
\end{equation}
where $Y(\omega)$ and $X(\omega)$ are the Fourier transforms of $Y_1^T$ and $X_1^T$ at frequency $\omega$, and $\varepsilon_{1}(\omega)$ and $\varepsilon_{2}(\omega)$ are the Fourier transforms of the errors of the models \eqref{VAR1.def} and \eqref{VAR2.def} at frequency $\omega$.
The components of the matrix are
\begin{equation*}
\vartheta_{lm}(\omega) = \delta_{lm} - \sum_{j=1}^p
\vartheta_{lm}(j)e^{(-i2{\pi}\omega j)},
\quad \text{ where } \quad
\left\{
\begin{array}{l}
\delta_{lm} = 0, \quad l = m, \\
\delta_{lm} = 1, \quad l \neq m,  
\end{array}, \quad l,m = 1,2. \right.
\end{equation*}
Rewriting equation \eqref{GC_frq1} as 
\begin{equation}
\left( \begin{array}{l} Y(\omega)\\ X(\omega) \end{array} \right) = 
\left( \begin{array}{ll} H_{11}(\omega) & H_{12}(\omega)\\ H_{21}(\omega) & H_{22}(\omega) \end{array} \right)
\left( \begin{array}{l} \varepsilon_{1}(\omega)\\ \varepsilon_{2}(\omega) \end{array} \right),
\label{GC_transferFUNCTION1}
\end{equation}
we have 
\begin{equation}
\left( \begin{array}{ll} H_{11}(\omega) & H_{12}(\omega)\\ H_{21}(\omega) & H_{22}(\omega) \end{array} \right) = 
\left( \begin{array}{ll} \vartheta_{11}(\omega) & \vartheta_{12}(\omega)\\ \vartheta_{21}(\omega) & \vartheta_{22}(\omega) \end{array} \right)^{-1},
\label{GC_transferFUNCTION2}
\end{equation}
where $H$ is the transfer matrix. The spectral matrix $S(\omega)$ can now be derived as 
\begin{equation}
S(\omega) =  H(\omega) \Sigma H^*(\omega), 
\end{equation}
where the asterisk denotes matrix transposition and complex conjugation. $\Sigma$ is the matrix defined in equation \eqref{SIGMA} (\cite{ding_granger_2006}). The spectral matrix $S(\omega)$ contains cross-spectra terms ($S_{12}(\omega)$, $S_{21}(\omega)$) and auto-spectra terms ($S_{11}(\omega)$, $S_{22}(\omega)$). If $X$ and $Y$ are independent, the cross-spectra terms are equal to zero.

Let us now write the auto-spectrum of $Y$ as
\begin{equation}
S(\omega)_{11} =  H(\omega)_{11} \Sigma_2 H^*(\omega)_{11}+2 \Gamma_{23} Re(H(\omega)_{11})H^*(\omega)_{12})+H(\omega)_{12}\Sigma_3H^*(\omega)_{12}. 
\label{AUTOSPECTRUM}
\end{equation}
In the following derivation, we will suppose that $\Gamma_{23}$, the off-diagonal element of the $\Sigma$ matrix in equation \eqref{SIGMA}, is equal to zero. In the case where this condition is not fulfilled, a more complex derivation is required (see \cite{ding_granger_2006} for further details).
If this independence condition is fulfilled, the auto-spectrum reduces to two terms,
\begin{equation}
S(\omega)_{11} =  H(\omega)_{11} \Sigma_2 H^*(\omega)_{11}+H(\omega)_{12}\Sigma_3H^*(\omega)_{12}. 
\label{AUTOSPECTRUM2}
\end{equation}
The first term, $H(\omega)_{11} \Sigma_2 H^*(\omega)_{11}$, only involves the variance of the signal of interest and thus can be viewed as the intrinsic part of the auto-spectrum. The second term $H(\omega)_{12}\Sigma_3H^*(\omega)_{12}$ only involves the variance of the second signal and thus can be viewed as the causal part of the auto-spectrum.

In Geweke's spectral formulation, the derivation of the spectral measure $f_{X \rightarrow{}Y}$ requires the fulfillment of several properties. The measures have to be non-negative, and the sum over all frequencies of the spectral Granger-causality components has to equal the time-domain Granger-causality quantity \eqref{F.def}:
\begin{equation}
\dfrac{1}{2\pi}\int\limits_{-\pi}^\pi{f_{X \rightarrow{}Y}(\omega)}d\omega=F_{X \rightarrow{}Y}.
\label{Geweke1.def}
\end{equation}
The two conditions together imply the desirable property
\begin{equation}
F_{X \rightarrow{}Y}=0 \Leftrightarrow f_{X \rightarrow{}Y}(\omega)=0, \quad \forall\omega.
\label{Geweke2.def}
\end{equation}
The third condition is that the spectral statistics have an empirical interpretation.
The spectral Granger-causality statistic proposed by Geweke fulfills all three requirements. For a given frequency $\omega$ and scalar variables $X$ and $Y$, it is defined as
\begin{equation}
f_{X \rightarrow{}Y}(\omega) = \dfrac{S_{11}(\omega)}{{H}_{11}(\omega)\Sigma_{2}{H}^{*}_{11}(\omega)},
\label{Geweke3.def}
\end{equation} 
where $\Sigma_{2}$ is the variance defined in equation~\eqref{VAR1.def}, $S_{11}(\omega)$ is the autospectrum of $Y$ and ${H}_{11}(\omega)$ is the $(1,1)$ element of the transfer matrix in equation \eqref{GC_transferFUNCTION2}.
The form of equation \eqref{Geweke3.def} provides an important interpretation: the causal influence depends on the relative size of the total power $S_{11}(\omega)$ and the intrinsic power ${H}_{11}(\omega)\Sigma_{2}{H}^{*}_{11}(\omega)$. Since the total power is the sum of the intrinsic and the causal powers (see equation~\eqref{AUTOSPECTRUM2}), the spectral Geweke--Granger-causality statistic is zero when the causal power is zero (i.e. when the intrinsic power equals the total power). The statistic increases as the causal power increases (\cite{ding_granger_2006}). 
Given the requirements imposed by Geweke, the measure $f_{X \rightarrow{}Y}(\omega)$ has a clear interpretation: it represents the portion of the power spectrum associated with the innovation process of model \eqref{VAR1.def}. However, this interpretation relies on the VAR model because the innovation process is only well-defined in this context (see \cite{brovelli_beta_2004}, \cite{chen_dynamics_2009},  \cite{chen_frequency_2006} and  \cite{bressler_cortical_2007} for examples of application in neuroscience).

The estimation of the parameters and the model order selection procedure is the same as in Section \ref{Granger causality criterion based on coefficients}, because the frequency-domain VAR model in equation \eqref{GC_frq1} is directly derived from the time-domain VAR model. The model order selection has to be performed within the time-domain model estimation procedure (see \cite{brovelli_beta_2004} and \cite{lin_dynamic_2009}).

In \cite{lin_dynamic_2009}, authors showed that under the null hypothesis $f_{X \rightarrow Y}(\omega) =0$ and based on \eqref{Geweke3.def}, one can derive a statistic that follows an $F$ distribution with degrees of freedom $(p, T-2p)$ when the number of observations tends to infinity (it was first derived in \cite{brovelli_beta_2004} and \cite{gourevitch_linear_2006}).

\subsection{Directed transfer function and partial directed coherence} 
\label{DTFpdc}
The directed transfer function (DTF) and the partial directed coherence (PDC) are alternative measures also derived from VAR estimated quantities that are closely related to the Geweke--Granger-causality statistic.

The DTF is a frequency-domain measure of causal influence based on the elements of the transfer matrix $H(\omega)$ in equation \eqref{GC_transferFUNCTION2}. It has both normalized (\cite{kaminski_evaluating_2001}) and non-normalized (\cite{kaminski_multichannel_2007}) forms.
The PDC (\cite{baccala_partial_2001}) is derived from the  matrix of the Fourier-transformation of the estimated VAR coefficients in equation \eqref{GC_frq1}. It provides a test for non-zero coefficients of this matrix. See \cite{schelter_assessing_2009} for a renormalized version of PDC and \cite{schelter_testing_2006} for an example of application in neuroscience. 

The DTF is expressed as
\begin{equation}
\text{DTF}_{X \rightarrow{}Y}(\omega)=\sqrt{\dfrac{|H_{12}(\omega)|^{2}}{|H_{11}(\omega)|^{2}+|H_{12}(\omega)|^{2}}},
\label{DTF}
\end{equation}
where $H_{12}(\omega)$ is the element $(1,2)$ of the transfer matrix in equation  \eqref{GC_transferFUNCTION2}. The PDC is defined as
\begin{equation}
\text{PDC}_{X \rightarrow{}Y}(\omega)=\dfrac{{\vartheta}_{12}(\omega)}{\boldsymbol{\vartheta}_{2}^{*}(\omega)\boldsymbol{\vartheta}_{2}(\omega)}, 
\label{PDC}
\end{equation}
where $\vartheta_{12}(\omega)$ represents the Fourier transformed VAR coefficient (i.e. the causal influence from $X$ to $Y$ at frequency $\omega$), and $ \boldsymbol{\vartheta}_{2}(\omega)$ represents all outflows from $X$.

The PDC is normalized, but in a different way from the DTF. Indeed, the PDC represents the outflow from $X$ to $Y$, normalized by the total amount of outflows from $X$. The  normalized DTF however represents the inflow from $X$ to $Y$, normalized by the total amount of inflows to $Y$. 

Comparisons between the Geweke--Granger-causality statistic, the DTF and the PDC are discussed  in \cite{eichler_evaluation_2006}, \cite{baccala_partial_2001}, \cite{gourevitch_linear_2006}, 
\cite{pereda_nonlinear_2005}, \cite{winterhalder_comparison_2005}, \cite{winterhalder_detection_2006} and more recently in the context of information theory in \cite{chicharro_spectral_2011}. As shown in \cite{chicharro_spectral_2011}, the causal interpretation of the DTF and the GGC, at least in the bivariate case, relies on Granger's definition of causality. For the PDC, a causal interpretation is different, as it relies on Sim's definition of causality (\cite{sims_money_1972}). See \cite{chamberlain_general_1982} and \cite{kursteiner} for a global overview and comparison of these two definitions of causality.
Finally, \cite{winterhalder_comparison_2005} conducted a simulation-based comparison of the DTF and the PDC (and other statistics) in a neuroscience context.

Unlike the original time-domain formulation of Granger causality,  statistical properties of these spectral measures have yet to be fully elucidated. For instance, the influence of signal pre-processing (e.g., smoothing, filtering) is not well established.

\subsubsection{Assessment of significance}
Theoretical distributions for DTF and PDC have been derived and are listed below. They are all based on the asymptotic normality of the estimated VAR coefficients. Therefore, they can be used and interpreted only if the assumptions behind this model hold.
\cite{schelter_testing_2006} showed that the PDC statistic asymptotically follows a $\chi^{2}$ distribution with $1$ degree of freedom. Furthermore, \cite{schelter_assessing_2009} showed that a  renormalized form of PDC can be related to a $\chi^{2}$ distribution with $2$ degrees of freedom. Finally, \cite{winterhalder_comparison_2005} provide simulations that suggest  that this $\chi^{2}$ distribution even works well if the true model order is strongly overestimated.
 
\cite{eichler_evaluation_2006} showed that the DTF quantity can be compared to a $\chi^{2}$ distribution with $1$ degree of freedom. This property is also based on the asymptotic normality of estimated VAR coefficients and its accuracy is evaluated through simulations.

For the PDC as well as for the DTF asymptotic distributions, \cite{schelter_b._quantification_2005} and \cite{eichler_evaluation_2006} state that a major drawback is that there are a lot of tests -- one for each frequency.
It is well known that when many tests are produced, caution has to be taken in interpreting those that are significant. For example, even under the null hypothesis of no information flow, there is a high probability that for a few frequencies the test will be significant.

\section{Time-Varying Granger Causality}
\label{Time-varying Granger causality}

Neuroscience data are nonstationary in most cases. The specificity (task or stimulus
related) of the increase or decrease and/or
local field potential implies this nonstationarity
which is of primary interest. A Granger-causality statistic
in the time- or the frequency-domain is desirable as it would capture the
evolution of Granger causality through time.

Since the original statistics are based on AR and VAR
models, and therefore on assumptions assuming that the autocorelation does not
vary along the time,
these models have to be extended to cases assuming changing autocorelation structure in order to suitably extract a Granger-causality statistic.

Practically, getting a statistic to assess the causality between two series for each time requires the estimation of the densities $f_t(Y_{t}|Y_{t-1}^{t-p})$ and $f_t(Y_{t}|Y_{t-1}^{t-p},X_{t-1}^{t-p})$ separately for each time $t$.
There are two additional difficulties to keep in mind. The first is the necessity of an objective criterion for time-varying model order selection and the second is the difficulty of incorporating all the recorded data (meaning all the trials) in the estimation procedure.

\subsection{Non-parametric statistics}
\subsubsection{Wavelet-based statistic} 
In the context of neuroscience, \cite{dhamala_analyzing_2008} proposed to bypass the nonstationarity problem by non-parametrically estimating the quantities which allow us to derive the spectral Geweke--Granger-causality (GGC) statistic \eqref{Geweke3.def}.
They derived an evolutionary spectral density through the continuous wavelet transform of the data, and then derived a quantity related to the transfer function (by spectral matrix factorization). Based on this quantity, they obtain a GGC statistic that can be interpreted as a time-varying version of the GGC statistic defined in \eqref{Geweke3.def}.

This approach bypassed the delicate step of estimating $f_t(Y_{t}|Y_{t-1}^{t-p})$ and $f_t(Y_{t}|Y_{t-1}^{t-p},X_{t-1}^{t-p})$ separately for each time. However this method presents several drawbacks in terms of interpretation of the resulting quantity. The GGC statistic is indeed derived from a VAR model and its interpretation directly follows from the causal nature of the  VAR coefficients. The non-parametric wavelet spectral density however does not have this Granger-causality interpretation. Therefore attention must be paid when interpreting this proposed evolutionary causal GGC statistic derived from spectral quantities which are not based on a VAR model.

\subsubsection{Local transfer entropy}
\cite{lizier_local_2008, lizier_multivariate_2011} and \cite{prokopenko_thermodynamic_2013} proposed a time-varying version of the transfer entropy \eqref{TR.def}, in order to detect dynamical causal structure in a functional magnetic resonance imaging (FMRI) study context.
The ``global'' transfer entropy defined in equation \eqref{TR.def} can be expressed as a sum of ``local transfer entropies'' at each time:
\begin{equation}
T_{X \rightarrow{}Y}= \dfrac{1}{T} \displaystyle\sum\limits_{t=1}^T f_t(y_{t}|y_{t-1}^{t-p},x_{t-1}^{t-p})\ln\dfrac{f_t(y_{t}|y_{t-1}^{t-p},x_{t-1}^{t-p})}{f_t(y_{t}|y_{t-1}^{t-p})}, 
\label{TR_local1.def}
\end{equation}
where each summed quantity can be interpreted as a single ``local transfer entropy'':
\begin{equation}
t_{x \rightarrow{}y}(t)= \ln\dfrac{f_t(y_{t}|y_{t-1}^{t-p},x_{t-1}^{t-p})}{f_t(y_{t}|y_{t-1}^{t-p})}.
\label{TR_local2.def}
\end{equation}
The step from equation \eqref{TR.def} to equation \eqref{TR_local1.def} is obtained by replacing the joint density $f(Y_{t},Y_{t-1}^{t-p},X_{t-1}^{t-p})$ with its empirical version. This simplification seems difficult to justify in a neuroscience context, considering the continuous nature of the data. In fact, the sampling rate used in neuroscience data acquisition is often very high. As such, this local transfer entropy does not seem to be a suitable time-varying causality statistic for an application in neuroscience. 
Moreover, even if the overall quantity in equation \eqref{TR.def} can be suitably expressed 
as a sum of orthogonal parts as in equation \eqref{TR_local2.def}, its causal nature does not necessarily remain in each part. As such, we cannot directly interpret these parts as causal, even if the sum of them gives an overall quantity that has an intrinsic causal meaning. Finally, \cite{prokopenko_thermodynamic_2013} or \cite{lizier_local_2008,lizier_multivariate_2011} do not provide an objective criterion for model order selection.

\subsection{Time-varying VAR model}
As seen before in equations \eqref{Fstat.def}, \eqref{WALD.def}, \eqref{Geweke3.def}, \eqref{DTF} and in \eqref{PDC}, parametric Granger-causality statistics in the time- and frequency-domains are derived from AR and VAR modelling of the data (equations \eqref{AR.def} and \eqref{VAR1.def} respectively). One way to extend these statistics to the nonstationary case amounts to allowing the AR and VAR parameters to evolve in time.
In addition to the difficulties related to model order selection and the fact that we have to deal with several trials, time-varying AR and VAR models are difficult to estimate since the number of parameters is most of the time considerable compared to the available number of observations. To overcome the dimensionality of this problem, 
\cite{chen_vector_2012} propose to make one of the three following assumptions, local stationarity of the process (\cite{dahlhaus_fitting_1997}), slowly-varying nonstationary characteristics (\cite{priestley_evolutionary_1965})
and slowly varying parameters for nonstationary models (\cite{ledolter_recursive_1980}).
In practice, it is difficult to distinguish between these assumptions but they all allow nonstationarity.
\cite{chen_vector_2012}  asserts that if one of the above assumptions is fulfilled, the estimate of a signal at some specific time can be approximated and inferred using the neighbourhood of this time point. 
Probably all time-varying methods proposed in the literature are based on one of these characteristics. 

We will discuss now the two widely-used approaches that deal with this type of nonstationarity: the windowing approach, based on the locally stationary assumption and the adaptive estimation approach, based on slowly-varying parameters.

\subsubsection{Windowing approach} 
A classical approach to adapt VAR models to the nonstationary case is windowing. 
This methodology consists in estimating VAR models in short temporal sliding windows where the underlying process is assumed to be (locally) stationary. See \cite{ding_short-window_2000} for a methodological tutorial on windowing estimate in neuroscience and \cite{long_nonstationary_2005} and \cite{hoerzer_directed_2010} for some applications in neuroscience.

The segment or window length is a trade-off between the accuracy of the parameters estimates and the resolution in time. The shorter the segment length, the higher the time resolution but also the larger the variance of the estimated coefficients. The choice of the model order is a related very important issue. With a short segment, the model order is limited, especially since we do not have enough residuals to check the quality of the fit in each window. Some criteria have been proposed in order to simultaneously optimize the window length and model order (\cite{lin_dynamic_2009,long_nonstationary_2005,solo_signal_2001}). This windowing methodology was extensively analyzed and commented in \cite{cekic_lien_2010}.
This method can easily incorporate several recorded trials in the analysis by combining all of them for the parameter estimate (\cite{ding_short-window_2000}). 

In \cite{cekic_lien_2010}, we found that this windowing methodology has several limitations. First, increasing the time resolution implies short time windows and thus too few residuals to assess the quality of the fit. Second, the size of the temporal windows is somehow subjective (even if it depends on a criterion), as is the overlap between the time windows. The order of the model in turn depends on the size of the windows and so the quality of the estimate strongly relies on several subjective parameters.

\subsubsection{Adaptive estimation method} 
A second existing methodology for estimating time-varying AR and VAR models is adaptive algorithms.
They consist in estimating a different model at each time, and not inside overlapped time windows.
The principle is always the same: the observations at time $t$ are expressed as a linear combination of the past values with coefficients evolving slowly over time plus an error term. The difference between the methods lies in the form of transition and update from coefficients at time $t$ to those at time $t+1$. This transition is always based on the prediction error at time $t$ (see \cite{schlogl_electroencephalogram_2000}).
The scheme is 
\begin{equation}
 \left\{ \begin{aligned}
    \boldsymbol{\varphi}_{t+1}&=&f(\boldsymbol{\varphi}_t,\boldsymbol{w}_t) \\
    Z_t &=&C_t \boldsymbol{\varphi}_t+\boldsymbol{v}_t      
       \end{aligned}
 \right.
 \qquad \text{with}\qquad
 \begin{cases}
    \boldsymbol{\varphi}_t&= \vect{[\boldsymbol{\vartheta}_{1(t)} ,\boldsymbol{\vartheta}_{2(t)},..,\boldsymbol{\vartheta}_{p(t)}]}', \\
    Z_t&=\begin{pmatrix}Y_t\\X_t\end{pmatrix},\\
    C_t \boldsymbol{\varphi}_t&=\sum\limits_{j=1}^p \vartheta_{j(t)} \begin{pmatrix}Y_t\\X_t\end{pmatrix},
    \end{cases}
\label{adaptive.def}
\end{equation}
where $\boldsymbol{\vartheta}_{j(t)}$ are the time-varying VAR coefficients at lag $j$ for time $t$, $\boldsymbol{v_t}$ is the error of the time-varying VAR equation at time $t$ and $\boldsymbol{w_t}$ is the error of the Markovian update of the time-varying VAR coefficients from time $t$ to time $t+1$.

There are several recursive algorithms to estimate this kind of model. 
They are based on the Least-Mean-Squares (LMS) approach (\cite{schack_parametrische_1993}) , the Recursive-Least-Squares (RLS) approach (see \cite{mainardi_pole-tracking_1995}, \cite{patomaki_tracking_1996}, \cite{patomaki_tracking_1995} and \cite{akay_biomedical_1994}  for basic developments, \cite{moller_instantaneous_2001} for an extension to multivariate and multi-trial data and \cite{astolfi_tracking_2008}, \cite{astolfi_time-varying_2010}, \cite{hesse_use_2003}, \cite{tarvainen_estimation_2004} and \cite{wilke_estimation_2008} for examples of application in neuroscience), and the Recursive AR (RAR) approach (\cite{bianchi_continuous_1997}). They are all described in detail in \cite{schlogl_electroencephalogram_2000}.

All these adaptive estimation methods depend on a free quantity that acts as a tuning parameter and defines the relative influence of $\boldsymbol{\varphi}_{t}$ and $\boldsymbol{w_t}$ on the recursive estimate of $\boldsymbol{\varphi}_{t+1}$. Generally this free tuning parameter determines the speed of adaptation, as well as the smoothness of the time-varying VAR parameter estimates. 
The sensitivity of the LMS, RLS and RAR algorithms to this tuning parameter was investigated in \cite{schlogl_electroencephalogram_2000} and estimation quality strongly depends on it. The ad-hoc nature of these procedures does not allow for proper statistical inference.

Finally, as for the previous models, the model order has to be selected. It is often optimized in terms of Mean Square Error, in parallel with tuning parameter selection (\cite{costa_adaptive_2011,schlogl_criterion_2000}).

\subsubsection{Kalman filter and the state space model}
\cite{kalman_new_1960} presented the original idea of the Kalman Filter. \cite{meinhold_understanding_1983} provided a Bayesian formulation.

A Kalman filtering algorithm can be used to estimate time-varying VAR models if it can be expressed in a state space form with the VAR parameters evolving in a Markovian way. This leads to the system of equations
\begin{equation}
 \left\{\begin{aligned}
 	\boldsymbol{\varphi}_{t+1}  =  A\boldsymbol{\varphi}_t+\boldsymbol{w}_t\qquad &\boldsymbol{w}_t\sim{N(0,Q)} \\
    Z_t  =  C_t \boldsymbol{\varphi}_t+\boldsymbol{v}_t \qquad &\boldsymbol{v}_t\sim{N(0,R)}\\ 
       \end{aligned}
 \right.
 \qquad \text{with}\qquad
 \begin{cases}
    \boldsymbol{\varphi}_t&=\vect{[\boldsymbol{\vartheta}_{1(t)} ,\boldsymbol{\vartheta}_{2(t)},\dots,\boldsymbol{\vartheta}_{p(t)}]}', \\
    Z_t&=\begin{pmatrix}Y_t\\X_t\end{pmatrix},\\
    C_t \boldsymbol{\varphi}_t&=\sum\limits_{j=1}^p \vartheta_{j(t)} \begin{pmatrix}Y_t\\X_t\end{pmatrix},
    \end{cases}
 \label{kalman_eq.def}
\end{equation}
where the vector $\boldsymbol{\varphi}_t$ contains the time-varying VAR coefficients that are adaptively estimated through the Kalman filter equations. The matrix $Q$ represents the variance-covariance matrix of the state equation that defines the Markovian process of the time-varying VAR coefficients. The matrix $R$ is the variance-covariance matrix of the observed equation containing the time-varying VAR model equation. 

With known parameters $A$, $Q$ and $R$, the Kalman smoother algorithm gives the best linear unbiased estimator for the state vector (\cite{kalman_new_1960}), which here contains the time-varying VAR coefficients of interest.

In the engineering and neuroscience literature, the matrix $A$ is systematically chosen as the identity matrix and $Q$ and $R$ are often estimated through some ad-hoc estimation procedures. These procedures and their relative references are listed in Tables \ref{Table 1} and \ref{Table 2}, which are based on \cite{schlogl_electroencephalogram_2000}.

There are many applications of these estimation procedures in the neuroscience literature, see for example \cite{vicente_transfer_2011}, \cite{roebroeck_mapping_2005}, \cite{hesse_use_2003}, \cite{moller_instantaneous_2001}, \cite{astolfi_tracking_2008}, \cite{astolfi_time-varying_2010} and \cite{arnold_adaptive_1998}. For an extension to several trials, the reader is referred to \cite{milde_dynamics_2011,milde_new_2010} and to \cite{havlicek_dynamic_2010} for an extension to forward and backward filter estimation procedure.

Any given method must provide a way to estimate the parameter matrices $A$, $Q$, and $R$ simultaneously with the state vector $\boldsymbol{\varphi}_{t+1}$, while selecting the model order in a suitable way. The procedure must also manage models based on several trials.

In the statistics literature, it has been known for a long time that the matrices $A$, $Q$, and $R$ can be obtained through a maximum likelihood EM-based approach (see \cite{shumway_approach_1982} and \cite{cassidy_bayesian_2002} for a Bayesian extension of this methodology).

\subsubsection{Wavelet dynamic vector autoregressive model} 
\label{Wavelet dynamic vector autoregressive model}

In order to derive a dynamic Granger-causality statistic in an FMRI experiment context, \cite{sato_method_2006} proposed another time-varying VAR model estimation procedure based on a wavelet expansion. 
They allow a time-varying structure for the VAR coefficients as well as for the variance-covariance matrix, in a linear Gaussian context. Their model is expressed as
\begin{equation}
f_t(Y_{t}|Y_{t-1}^{t-p},X_{t-1}^{t-p})  =  \phi \big(Y_{t}; \mu=\displaystyle\sum\limits_{j=1}^p
\vartheta_{11(j)}(t) Y_{t-j}+\displaystyle\sum\limits_{j=1}^p \vartheta_{12(j)}(t) X_{t-j}, \sigma(t)^2=\Sigma(t) \big),
\end{equation}
where $\vartheta_{11(j)}(t)$ and $\vartheta_{12(j)}(t)$ are the time-varying VAR coefficients at time $t$ and $\Sigma(t)$ is the time-varying variance-covariance matrix at time $t$. These are both unknown quantities that have to be estimated.

They make use of the wavelet expansion of functions in order to estimate the time-varying VAR coefficients and the time-varying variance-covariance matrix. As any function can be expressed as a linear combination of wavelet functions, \cite{sato_method_2006} consider the dynamic VAR coefficient vector $\boldsymbol{\vartheta}(t)$ and the dynamic covariance matrix $\Sigma_{t}$ as functions of time, and so expressed them as a linear combination of wavelet functions.

They proposed a two-step iterative generalized least square estimation procedure. The first step consists in estimating the coefficients of the expanded wavelet functions using a generalized least squares procedure. In the second step, the squared residuals obtained in the previous step are used to estimate the 
wavelet expansion functions for the covariance  matrix  $\Sigma_{t}$ (see \cite{sato_method_2006} for further details).

The authors gave asymptotic properties for the parameter estimates, and statistical assessment of Granger-causal connectivities is achieved through a time-varying Wald-type statistic as described in equation \eqref{WALD.def}. An application in the context of gene expression regulatory network modelling can be found in 
\cite{fujita_time-varying_2007}.

This wavelet-based dynamic VAR model estimation methodology has the advantage of avoiding both stationarity and linearity assumptions. However there is, surprisingly, no mention of a model order selection criterion and the question how to take into account all the recorded trials in the estimation procedure is not addressed.

\section{Existing Toolboxes} 
\label{Existing toolboxes}
Several toolboxes to analyse neuroscience data have been made available in recent years. We will only list those providing estimate of time-varying VAR models and Granger-causality statistics. Tables \ref{Table 3a} and \ref{Table 3b} present a list of these toolboxes, with references and details of their content. The description of the content is not exhaustive and all of them contain utilities beyond (time-varying) VAR model estimate and Granger-causality analysis.

\section{Discussion}
\label{Discussion} 

\subsection{Limitations}

This article does not discuss symmetric functional connectivity statistics such as correlation and coherence. The reader is referred to \cite{delorme_eeglab_2011} and \cite{pereda_nonlinear_2005} for an overall review of these statistics in the time and frequency domains. This symmetric connectivity aspect is also very important and carries a lot of information but its presentation is beyond the scope of this article which propose a review of all existing methods allowing us to derive a time-varying Granger-causality statistic.

We do not discuss other existing tools to analyse effective connectivities either. The most popular is certainly the dynamic causal modelling (DCM) of \cite{friston_functional_1994} and \cite{friston_dynamic_2003}, which is based on nonlinear input-state-output systems and bilinear approximation of dynamic interactions. DCM results strongly rely on prior connectivity specifications and especially on the assumption of stationarity. Therefore the lack of reference to the DCM methodology here is explained by its unsuitability in the context of nonstationarity.

Another important topic not highlighted here is the estimation procedure and interpretation of Granger-causality statistics in a multivariate context. As discussed in Section \ref{DTFpdc},  by their relative normalization, the DTF and PDC statistics take into account the influence of other information flows when testing for a causal relationship between two signals.
Another measure is conditional Granger causality, which was briefly mentioned in equation \eqref{multiple_GC.def}. Indeed when three or more simultaneous brain areas are recorded, the causal relation between any two of the series may either be direct, or be mediated by a third, or a combination of both. These cases can be addressed by conditional Granger causality, which has the ability to determine whether the interaction between two time series is direct or mediated by another one. Conditional Granger causality in time- and frequency-domains is described in \cite{ding_granger_2006}, based on previous work of \cite{geweke_measures_1984}.

Finally, an important extension is partial Granger causality. As described in \cite{bressler_wienergranger_2011} and \cite{seth2010matlab}, all brain connectivity analyses involve variable selection, in which the relevant set of recording brain regions is selected for the analysis. In practice, this step may exclude some relevant variables. The lack of exogenous and latent inputs in the model can lead to the detection of apparent causal interactions that are actually spurious. The response of \cite{guo_partial_2008} to this challenge is what is called partial Granger causality. This is based on the same intuition as partial coherence, namely that the influence of exogenous and/or latent variables on a recorded system will be highlighted by the correlations among residuals of the VAR modelling of the selected measured variables. \cite{guo_partial_2008} also provide an extension in the frequency domain.

\subsection{EEG and fMRI application} 

The application of Granger-causality methods to FMRI data is very promising, given the high spatial resolution of the FMRI BOLD signal, as shown in \cite{bressler_wienergranger_2011} and \cite{seth2010matlab}.

However FMRI data are subject to several potential artifacts, which complicates the application of Granger-causality methods to these specific data (\cite{roebroeck_mapping_2005}). These potential artifacts come from the relatively poor temporal resolution of the FMRI BOLD signal, and from the fact that it is an indirect measure of neural activity. This indirect measure is usually modelled by a convolution of this underlying activity with the hemodynamic response function (HRF). A particularly important issue is that the delay of the HRF is known to vary between individuals and between different brain regions of the same subject, which is an important issue given that Granger causality is based on temporal precedence.
Furthermore, several findings indicate that the BOLD signal might be biased for specific kinds of neuronal activities (e.g., higher BOLD response for gamma range compared to lower frequencies, \cite{niessing_hemodynamic_2005}). The impact of HRF on Granger-causality analysis in the context of BOLD signals has recently been discussed in \cite{roebroeck_identification_2011}.

The very high time resolution offered by magnetoencephalography (MEG) or electroencephalography (EEG) methods on the surface or during intracranial recordings allows the application of Granger-causality analysis to these data to be very powerful (\cite{bressler_wienergranger_2011}). An application of spectral Granger-causality statistics for discovering causal relationships at different frequencies in MEG and EEG data can be found for example in \cite{astolfi_comparison_2007}, \cite{bressler_cortical_2007} and \cite{brovelli_beta_2004}.
A key problem with the application of Granger-causality methods to MEG and EEG data is the introduction of causal artifacts during preprocessing. Bandpass filtering for example can cause severe confounding in Granger-causality analysis by introducing temporal correlations in MEG and EEG time series (\cite{seth2010matlab,florin_effect_2010}).

The reader is referred to \cite{bressler_wienergranger_2011} and \cite{seth2010matlab} for a thorough discussion of the application of Granger-causality methods to fMRI, EEG and MEG data.

\subsection{Neuroscience data specificities} 

As described  in \cite{vicente_transfer_2011}, neuroscience data have specific characteristics which complicates effective connectivity analysis. For example, the causal interaction may not be instantaneous but delayed over a certain time interval $(\upsilon)$, so the history of the variables $Y$ and $X$ in equation \eqref{VAR1.def} has to be taken from time $t-\upsilon-1$ to $t-\upsilon-p$, instead of from time $t-1$ to $t-p$, depending on the research hypothesis.  

Another very important parameter to choose is the time-lag $\tau$ between the data points in the history of $Y$ and $X$, which permits more parsimonious models. Choosing a certain time-lag parameter means that the causal history of variables $Y$ and $X$ should be selected by taking the time-points from $t-\upsilon-1$ to $t-\upsilon-\tau p$, all of them being spaced by a lag $\tau$.
This is a very useful tool for dealing with high or low frequency modulations of the data, as high frequency phenomena needs a small time-lag and conversely for low frequency phenomena.

This time-lag parameter $\tau$ has a clear and interpretable influence on Granger-causality statistics in the time-domain, which directly relies on the estimated VAR parameters. It is however very difficult to see what its impact is on the frequency-domain causality-statistics, where the time-domain parameter estimates are Fourier transformed and only then interpreted as a causality measure at each frequency. 

\subsection{Asymptotic distributions} 

As we have seen in Sections \ref{Time domain causality} and \ref{Frequency domain causality}, time-domain Granger-causality statistics in equations \eqref{Fstat.def} and \eqref{WALD.def} asymptotically follow $F$ and $\chi^2$ distributions. Frequency-domain causality statistics in equations \eqref{DTF} and \eqref{PDC} are both asymptotically related to a $\chi^{2}$ distribution. ``Asymptotic'' here means when the number of observations $T$ goes to infinity.

These distributions have the advantage of requiring very little computational time compared to bootstrap or permutation surrogate statistics. However, one has to be aware that all these properties are derived from the asymptotic properties of the VAR estimated coefficients. They are thus accurate only if the assumptions behind VAR modelling are fulfilled. They also may be very approximate when the number of sample points is not large enough. 

Since in neuroscience causal hypotheses are often numerous (in terms of number of channels or/and number of specific hypothesis to test), these distributions can nonetheless provide a very useful tool allowing us to rapidly check for statistical significance of several causality hypotheses. They thus offer a quick overview of the overall causal relationships.

An important aspect is that the tests based either on the asymptotic distributions or on resampling are only pointwise significance tests. Therefore, when jointly testing a collection of values for a complete time or frequency or time-frequency connectivity map, it is important to suitably correct the significance threshold for multiple comparisons.

\section{Conclusion}

Neuroscience hypotheses are often relatively complex, such as asking about time-varying causal relationships specific to certain frequency bands and even sometimes between different frequency bands (so-called cross-frequency coupling).

Granger causality is a promising statistical tool for dealing with some of  these complicated research questions about effective connectivity. However the postulated models behind have to be suitably estimated in order to derive accurate statistics.

In this article we have reviewed and described existing Granger-causality statistics and focused on model estimation methods that possess a time-varying extension. Time-varying Granger causality is of primary interest in neuroscience since recorded data are intrinsically nonstationary. However, its implementation is not trivial as it depends on the complex estimate of time-varying densities.
We reviewed existing methods providing time-varying Granger-causality statistics and discussed their qualities, limits and drawbacks.

\begin{table}[!htb]
\centering
\setcellgapes{1pt}
\makegapedcells
\newcolumntype{L}[1]{>{\raggedright\arraybackslash }b{#1}}
\newcolumntype{C}[1]{>{\centering\arraybackslash }b{#1}}
\begin{tabular}{C{2.5cm}|L{8cm}|C{2.5cm}}
Type&Estimate of $R_t$& References \\
\hline 
Univariate  & $R_t = (1-UC)R_{t-1}  + UC {e_t}^2$& (\cite{schack_parametrische_1993}) \\
One trial& $e_t=y_t-C_tx_t$&\\
\hline
Multivariate & $R_0=I_d$ & (\cite{milde_new_2010})\\
Multiple trial& $\overline{R_t}=\overline{R_{t-1}}(1-UC)  + UC {e}' e/(K-1)$& \\\hline
Univariate & $R_t =1$& (\cite{isaksson_computer_1981}) \\
One trial&&
\\\hline
Univariate& $R_t =1-UC$& (\cite{patomaki_tracking_1996})\\
One trial && (\cite{patomaki_tracking_1995})\\
&& (\cite{haykin_adaptive_1997})\\
&& (\cite{akay_biomedical_1994}) 
\\\hline
Univariate& $q_t = {Y_{t-1}}' A_{t-1} Y_{t-1}$& (\cite{jazwinski_adaptive_1969}) \\
One trial & ${R_t}^+ = \left\{ \begin{array}{lll}   (1-UC){R_{t-1}}^+ +UC(e_t- q_t) &  \text{if} &
  {e_t}^2 > q_t \\
   {R_{t-1}}^+ & \text{if} & {e_t}^2 \leq q_t \end{array} \right. $ &\\
&$R_t={R_t}^+$&
\\\hline 
Univariate& Same as \cite{jazwinski_adaptive_1969} except& (\cite{penny_dynamic_1998}) \\
One trial & $R_t={R_{t-1}}^+$& 
\\\hline
Univariate& $R_t  = 0$& (\cite{kalman_new_1960})\\ 
One trial & & (\cite{kalman_new_1961})
\\\hline
\end{tabular}
\caption{Variants for estimating the covariance matrix $R_t$ based on \cite{schlogl_electroencephalogram_2000}.
$UC$ acts as tuning parameters that has to be choosing between $0$ and $1$.}
\label{Table 1}
\end{table}

\begin{table}[!htb]
\centering
\setcellgapes{1pt}
\makegapedcells
\newcolumntype{L}[1]{>{\raggedright\arraybackslash }b{#1}}
\newcolumntype{C}[1]{>{\centering\arraybackslash }b{#1}}
\begin{tabular}{C{3cm}|L{7cm}|C{3cm}}
    Type&Estimate of $Q_t$& References 
    \\ \hline
    Univariate & $Q_t  = UCx_t$& (\cite{akay_biomedical_1994}) \\
    One trial&& (\cite{haykin_adaptive_1997})
    \\\hline
    Univariate& $x_t  =(I-k_t ){y_{t-1}}' A_{t-1}$ & (\cite{isaksson_computer_1981})\\
    One trial& $Q_t  ={UC}^2I$& 
    \\\hline
    Univariate& $K_t  ={y_{t-1}}' x_{t-1}{y_{t-1}}'+R_t$& (\cite{jazwinski_adaptive_1969}) \\  
    One trial & $L_t  =(1-UC)L_{t-1}+\dfrac{UC*({e_t}^2-K_t)}{{y_{t-1}}' y_{t-1}}$ & (\cite{penny_dynamic_1998})\\
    & $Q_t  =  \left\{ \begin{array}{lll} L_tI &  \text{if} & L_t>0 \\
           0 & \text{if} &  L_t \leq 0
\end{array} \right.$ & 		\\\hline 
\end{tabular}
\caption{Variants for estimating the covariance matrix $Q_t$ based on \cite{schlogl_electroencephalogram_2000}.}
\label{Table 2}
\end{table}

\begin{landscape}

\begin{table}[!htb]
\centering
\setcellgapes{1pt}
\makegapedcells
\newcolumntype{L}[1]{>{\raggedright\arraybackslash }b{#1}}
\newcolumntype{C}[1]{>{\centering\arraybackslash }b{#1}}

\begin{tabularx}{20.5cm}{p{2cm}|C{1.3cm}|p{8.5cm}|X}
    Toolbox &Software&TV-VAR implemented estimation method&Implemented statistics of causality 
    \\ \hline
    
    BSMART&Matlab& Windowing approach based on \cite{ding_short-window_2000} & Geweke-spectral Granger-causality statistic \eqref{Geweke3.def} \\  
    \footnotesize{Brain System for Multivariate AutoRegressive Time series}&& Implemented for single and multiple trials & \\   
    (\cite{cui_bsmart:_2008}) &&  & \\   
\\\hline

    BioSig &Matlab& Kalman filter estimation type (mvaar.m Matlab function)& No causality statistic implemented\\
   (\cite{schlogl_biosig_2008}) && Implemented for single trial only&  \\
   && Variants for estimating the covariance matrices $R_t$ and $Q_t$ are implemented based on \cite{schlogl_electroencephalogram_2000}&\\
    \\\hline
    
    GCCA &Matlab& Windowing approach based on \cite{ding_short-window_2000} & Geweke-spectral Granger-causality statistic \eqref{Geweke3.def}\\  
     \footnotesize{Granger Causal Connectivity Analysis} && Implemented for single and multiple trials & Partial Granger causality (\cite{guo_partial_2008,bressler_wienergranger_2011})\\ 
     (\cite{seth2010matlab})&& & Granger autonomy (\cite{bertschinger_autonomy:_2008,seth_measuring_2010})\\  
     &&  & Causal density (\cite{seth_causal_2005,seth_causal_2008}) \\  
	\\\hline 

\end{tabularx}
\caption{List of available toolboxes for estimating time-varying VAR models and Granger-causality statistics.}
\label{Table 3a}
\end{table}
\end{landscape}

\begin{landscape}

\begin{table}[!htb]
\centering
\setcellgapes{0.5pt}
\makegapedcells
\newcolumntype{L}[1]{>{\raggedright\arraybackslash }b{#1}}
\newcolumntype{C}[1]{>{\centering\arraybackslash }b{#1}}

\begin{tabularx}{20.5cm}{p{2.5cm}|C{1.3cm}|p{7.5cm}|X}
    Toolbox & Software & TV-VAR implemented estimation method&Implemented statistics of Causality 
    \\ \hline
 	
	eConnectome  & Matlab & Kalman filter estimation type (same mvaar.m Matlab function as BioSig toolbox)& Directed transfer function \eqref{DTF}   \\  
	  (\cite{he_econnectome:_2011})&& Implemented for single trial only& Adaptive version of directed transfer function (\cite{wilke_estimation_2008})\\
   	  && Variants for estimating the covariance matrices $R_t$ and $Q_t$ are implemented based on \cite{schlogl_electroencephalogram_2000}& \\
	\\\hline 
	
    SIFT  & Matlab& Windowing approach based on \cite{ding_short-window_2000} & Partial directed coherence \eqref{PDC} \\  
    \footnotesize{Source Information Flow Toolbox}  & & Implemented for single and multiple trials& Generalized partial directed coherence (\cite{baccala_generalized_2007}) \\  
    (\cite{delorme_eeglab_2011})&&  & Renormalized partial directed coherence (\cite{schelter_assessing_2009}) \\
    && Kalman filter estimation type (same mvaar.m Matlab function as BioSig toolbox) & Directed transfer function \eqref{DTF} \\
    && Implemented for single trial only &Full frequency directed transfer function (\cite{korzeniewska_determination_2003})\\
    &&  & Geweke--Granger-causality \eqref{Geweke3.def} \\

	\\\hline 
	
	GEDI & R &Wavelet dynamic vector autoregressive estimation method \ref{Wavelet dynamic vector autoregressive model}& Granger-causality criterion 2 \eqref{GC2a.def} and Wald statistic \eqref{WALD.def} (\cite{fujita_time-varying_2007}) \\  
	\footnotesize{Gene Expression Data Interpreter} & R &Wavelet dynamic vector autoregressive estimation method \ref{Wavelet dynamic vector autoregressive model}& \\  
	(\cite{fujita_time-varying_2007}) & &&  \\ 

	\\\hline 
\end{tabularx}
\caption{List of available toolboxes for estimating time-varying VAR models and Granger-causality statistics.}
\label{Table 3b}
\end{table}
\end{landscape}

\clearpage
\bibliographystyle{chicago}
\bibliography{stat_OR}

\end{document}